# 1.5 μm Epitaxially Regrown Photonic Crystal Surface Emitting Laser Diode

Zijun Bian, Katherine J. Rae, Adam F. McKenzie, Ben C. King, Nasser Babazadeh, Guangrui Li, Jonathan R. Orchard, Neil D. Gerrard, Stephen Thoms, Donald A. MacLaren, Richard J. E. Taylor, David Childs, and Richard A. Hogg, *Member, IEEE*

*Abstract*— We present an InP-based epitaxially regrown photonic crystal surface emitting laser diode, lasing in quasi-CW conditions at 1523nm.

## I. INTRODUCTION

There has been considerable recent interest in the development of photonic crystal surface emitting lasers [1,2]. A photonic crystal (PC) is formed through a 2D periodic variation in refractive index, and when placed within a laser structure, a surface emitting laser can be realised [3].

PCSELs have shown single mode operation [4,5], low divergence [6], polarisation and beam shape control [7-9], beam steering [10], high power and brightness [1], and coherently coupled arrays [11,12]. There has also been significant work on the simulation of devices [13-16] particularly to optimise output power through PC design [17-19], and more recently simulation results have suggested that speeds of 40 GHz could be achieved from a small, resonator embedded device [20].

Future 5G roll-out requires a corresponding increase in the size of data-centres, requiring longer link lengths, and due to problems in situating them in urban areas, to distributed data-centres, again requiring longer link-lengths approaching those of shorter haul metro networks. In order to service this growing data usage, reliable, low-cost, high speed, lasers operating at 1300 nm and 1550 nm will be required, which may be serviced by InP-based PCSEL devices. Additional applications for such surface emitting InP based devices include free-space communications, LiDAR, fibre-based sensing, *etc...*

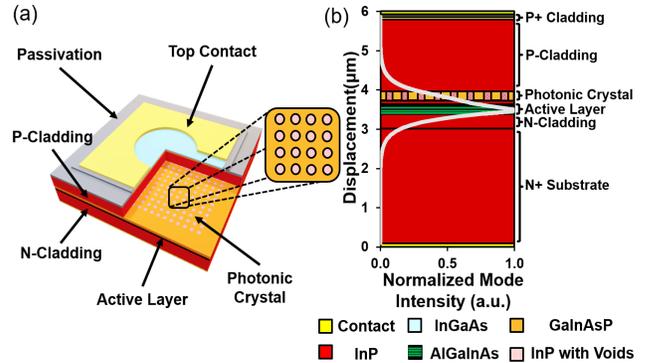

*Figure 1* a) A schematic of the device, with cut-out area showing the photonic crystal pattern and active layer. b) A schematic of the corresponding mode distribution inside the device structure. The epitaxial structure is shown, with the various materials in the key below.

PCSEL devices have traditionally been realised with wafer fusion [4-8], but more recently epitaxial overgrowth through MBE [21,22] and MOVPE [23-26] has been utilised. This move from wafer fusion to epitaxial regrowth has been critical for achieving high output powers [1,27]. This is because the wafer fusion interface contains regions of discontinuous crystallinity, that result in undesirable defect states. Epitaxial regrowth allows single-crystal PCSEL structures to be realised with encapsulated voids or fully in-filled photonic crystals [28].

In this paper we demonstrate an epitaxially re-grown InP PCSEL operating at 1523 nm at room temperature. The realization of epitaxially regrown PCSELs at this wavelength, as opposed to ones via wafer fusion [29], is a critical step in providing a route to engineer high-power sources with high beam quality.

This paper was submitted for review 7[th] April 2020. This work was supported by EP/L015323/1

A. F. McKenzie is supported by the Royal Commission for the Exhibition of 1851 and CST Global Ltd.

Z. Bian is with the James Watt School of Engineering, University of Glasgow, Glasgow, UK (e-mail: z.bian.1@research.gla.ac.uk).

K. J. Rae is with the James Watt School of Engineering, University of Glasgow, Glasgow, UK (e-mail: katherine.rae@glasgow.ac.uk).

A. F. McKenzie is jointly with the James Watt School of Engineering, University of Glasgow, Glasgow, UK, and CST Global Ltd. (e-mail: a.mckenzie.1@research.glasgow.ac.uk).

B. King is with the James Watt School of Engineering, University of Glasgow, Glasgow, UK (e-mail: Ben.king.1@research.glasgow.ac.uk).

N. Babazadeh was with the James Watt School of Engineering, University of Glasgow, Glasgow, UK (e-mail: n.babazadeh@sheffield.ac.uk).

G. Li is with the James Watt School of Engineering, University of Glasgow, Glasgow, UK (e-mail: guangrui.li@glasgow.ac.uk).

J. Orchard is with CST Global Ltd (e-mail: jorchard@cstglobal.uk).

N.D. Gerrard is with CST Global Ltd (e-mail: ngerrard@cstglobal.uk).

S. Thoms is with the James Watt School of Engineering, University of Glasgow, Glasgow, UK (e-mail: stephen.thoms@glasgow.ac.uk).

D.A. MacLaren is with the School of Physics and Astronomy, University of Glasgow, Glasgow, UK (e-mail: Donald.maclaren@glasgow.ac.uk).

R. J. E. Taylor is with the James Watt School of Engineering, University of Glasgow, Glasgow, UK (e-mail: richard.taylor@glasgow.ac.uk).

D. Childs is with the James Watt School of Engineering, University of Glasgow, Glasgow, UK (e-mail: david.childs@glasgow.ac.uk).

R. A. Hogg is with the James Watt School of Engineering, University of Glasgow, Glasgow, UK (e-mail: richard.hogg@glasgow.ac.uk).

## II. Device Design and Simulation

Figure 1 shows a schematic of our InP based epitaxially regrown PCSEL structure, where the structure consists of, from bottom to top, 3.3 μm of n-doped InP, five 6 nm AlGaInAs quantum wells (separated by 8 nm AlGaInAs barrier layers), a p-doped 243 nm PC layer (consisting of GaInAsP and InP with air containing voids), 1.8 μm of p-doped InP, and capped with a p+ InGaAs cap layer. The PC is a circular atom in a square lattice. Assuming an average refractive index for the PC layer of 3.24, the structure is simulated using the Ritz-Iteration method [30]. The mode shows an overlap with the QWs of 4.7% and overlap with the photonic crystal of 16.4 %.

Figure 2 (a) shows a conventional cross-sectional transmission electron microscope (TEM) image of our regrown photonic crystal grating layer (details later). The bright, central region corresponds to a crystallographic void formed within the etched feature; these are encapsulated laterally by GaInAsP and InP above. Such voids are observed within each of the grating holes. Due to the lattice-matched nature of the structure there is minimal contrast between the GaInAsP layer and adjacent InP layers. Figure 2 b) shows a schematic of the TEM image shown in Fig 2 a). The simulation of band structure is calculated by 2D plane-wave (PW) expansion method [31]. The dotted lines, denoted $n_{eff\_a}$ and $n_{eff\_b}$ show the cross section through the material stack used to calculate the refractive index of the atom, and field materials in the PC, hereafter referred to as region a and b, respectively [32].

Figure 3 shows the simulated photonic band structure for our PC calculated utilizing the structural information from Fig 2, where an r/a of 0.17 (where r/a is the ratio between radius and period of the photonic crystal) is determined, as is the position and height of the void. The k-vector is plotted to 0.2 (2π/λ), which was determined to be the collection angle of the optics used to measure the device (described later). As expected from the symmetry of the PC, four modes are obtained. A splitting of the bands at the Γ-point results in the high density of states from which lasing may occur. The insets in Fig 3 show band splitting of modes A and B. Shown adjacently are the in-plane electric field highlighting that bands C and D are leaky and bands A and B are non-leaky and therefor the likely lasing modes.

## III. Device Fabrication

The devices were fabricated on InP epitaxial wafers, designed to emit around 1550 nm. The epitaxial layer structure of the base planar was as follows (with doping levels in brackets); 3 μm - thick n+ InP buffer layer ($1.5 \times 10^{18}$ cm$^{-3}$), 300 nm n-type InP ($1 \times 10^{18}$ cm$^{-3}$), 15 nm n-type Al$_{0.43}$Ga$_{0.05}$In$_{0.53}$As ($1 \times 10^{18}$ cm$^{-3}$), 15 nm n-type Al$_{0.40}$Ga$_{0.08}$In$_{0.53}$As ($6 \times 10^{17}$ cm$^{-3}$), 12 nm un-doped Al$_{0.26}$Ga$_{0.25}$In$_{0.49}$As, five 60nm-thick Al$_{0.26}$Ga$_{0.25}$In$_{0.71}$As quantum wells with four 8nm – thick Al$_{0.26}$Ga$_{0.25}$In$_{0.49}$As barrier layers, 12 nm of un-doped Al$_{0.26}$Ga$_{0.25}$In$_{0.49}$As, 15 nm p-type Al$_{0.38}$Ga$_{0.09}$In$_{0.53}$As ($3 \times 10^{17}$ cm$^{-3}$), 90 nm p-type Al$_{0.45}$Ga$_{0.02}$In$_{0.53}$As ($4 \times 10^{17}$ cm$^{-3}$), 50 nm p-type InP cladding layer ($6 \times 10^{17}$ cm$^{-3}$), 50 nm p-type InP ($8 \times 10^{17}$ cm$^{-3}$) cladding layer, and a 243 nm - thick p+ Ga$_{0.22}$In$_{0.78}$As$_{0.48}$P$_{0.52}$ layer ($1 \times 10^{18}$ cm$^{-3}$).

On these base epitaxy structures, 200 nm of SiO$_2$ was deposited by plasma enhanced chemical vapour deposition. A square lattice, circular unit cell photonic crystal, with a period of 470 nm , was defined by electron-beam lithography in PMMA and etched into the SiO$_2$ by reactive ion etching with CHF$_3$/Ar chemistry. This acts as a hard mask for the etching of the underlying semiconductor. Through this hard mask, the semiconductor is etched to a depth of approximately 170 nm - just above the p-cladding layer atop the AlGaInAs active region with a CHF$_4$/ H$_2$ - based inductively coupled plasma etch. SEM images of test structures indicate that the etched InGaAsP PC layer is not modified by the regrowth process. See schematic in Fig 2 b).

The SiO$_2$ hard mask is then removed and epitaxial regrowth is undertaken. Immediately prior to regrowth, the wafer is uv/ozone cleaned (UVCOS uv/ozone cleaner) followed by 1 minute in 10:1 buffered HF. Regrowth was performed in an AIXTRON 2400 G2 Planetary MOVPE reactor at 100 mbar, utilising trimethylindium (TMIn) and trimethylgallium (TMGa) as group-III precursors, and phosphine (PH$_3$) and arsine (AsH$_3$) as group-V sources. Initially 243 nm of InP was deposited at a growth rate of 10 nm/min at 600 °C. Following this, 1.8 μm of p-doped InP was grown. Regrowth is finished with 25 nm of p-GaInAsP and a 75 nm - thick p+-InGaAs contact layer.

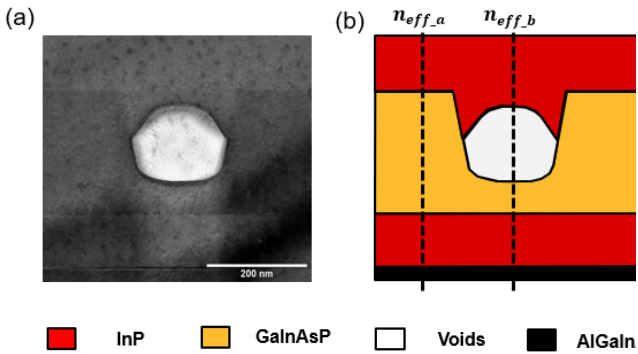

*Figure 2: a) A TEM image, and b) a schematic of one re-grown circular etched feature in the photonic crystal pattern. The etched GaInAsP can be seen, as well as the regrown InP that fills this etched feature, and a central air-void. The dotted lines in b) denote the cross section from which the effective refractive indices used in further simulations are taken.*

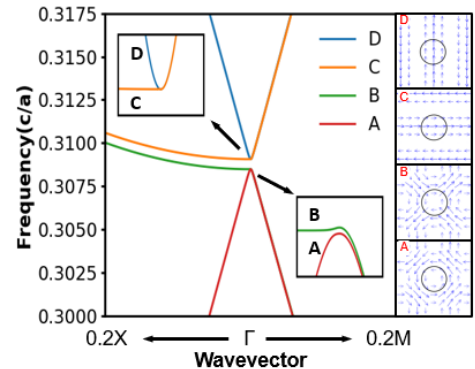

*Figure 3: a) The band structure around Γ-point band edges of a square lattice, circular unit cell photonic crystal, with insets showing degeneracy of the bands and the in plane electric field of each band shown adjacent, the black circles represent the edge of the refractive index contrast in the photonic crystal, and the arrow size represents the intensity of the electric field at each point*

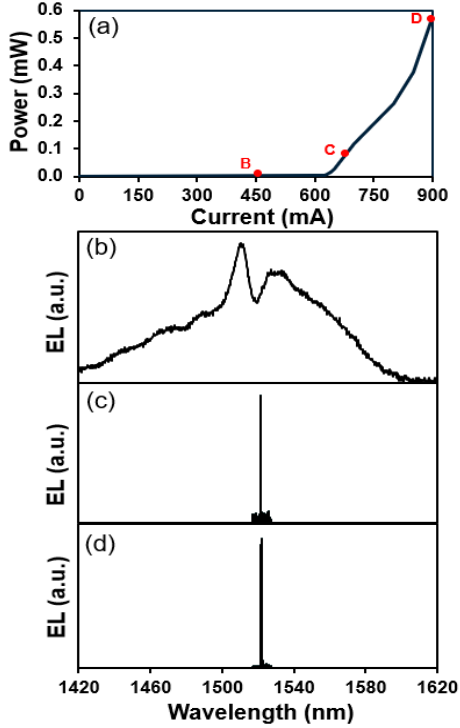

*Figure 4*: a) The current-power characteristics of an InP-based PCSEL, measured under pulsed conditions (10% duty cycle, 10 μs pulse width). b), c) and d) show the emission spectra of the device when driven at 450 mA, 700 mA, and 900 mA respectively.

Following re-growth, a 200 μm x 200 μm square mesa is patterned with photolithography and etched in a solution of sulphuric acid and hydrogen peroxide. This etches the regrown material, to a depth of 100 nm. A 200 nm - thick SiO$_2$ passivation layer is then deposited across the quarter wafer, with contact windows etched into this with a CHF$_3$ – and Ar – containing reactive ion etch. Following the opening of these contact windows, Ti/Pt/Au is deposited on the top surface of the quarter wafer. A 200 μm square contact with a "lollipop" shaped aperture of 100 nm diameter in the centre is created via a conventional metal lift-off process, the contact shape is illustrated in Fig 1 a). 57% of the device emission area is covered by this contact. A Ni/Au/Ge/Ni/Au n-type contact is then deposited on the bottom side of the quarter wafer, via electron-beam evaporation. This is then annealed at 400 °C for 1 minute. Thick Ti/Au bond-pads are added to the top surface of the quarter wafer, by an electron-beam evaporation and lift-off process.

IV. RESULTS

Devices were measured at 15 °C under quasi-cw conditions using a 10% duty cycle and 10μs pulse width. Surface emission was collected using a NA =0.34 lens, and was focused into a multi-mode fibre-optic cable. Electro-luminescence spectra were measured using an optical spectrum analyser with a resolution of 0.1 nm.

Figure 4 a) shows the current-power characteristics of a typical device showing a threshold current of 640 mA (J=1.6 kA/$cm^{-2}$). An average slope efficiency of ~0.002 W/A is obtained, which is low due to the circular symmetry of the PC [33,34] and the masking of the PCSEL emission by the contacts. Figure 4 b) shows the sub-threshold electro-luminescence at 450 mA. Two main peaks are observed at 1512 nm and 1527 nm, attributed to the modification of the spontaneous emission spectrum by the PC. Figures 4 c) and 4 d) show the electroluminescence spectra of the same device at 700 and 900 mA, respectively. A clear lasing peak is observed at 1523 nm in both cases. Figure 5a) shows the EL spectrum of the PCSEL in Fig 4 c) plotted over a narrower wavelength range.

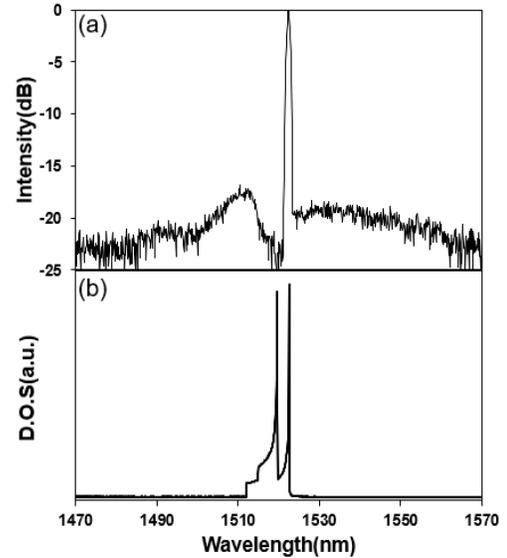

*Figure 5:* a) The emission spectrum of a typical device, plotted from 1470 nm to 1570 nm. b) The corresponding simulated optical density of states.

Figure 5 b) shows the simulated optical density of states (ODOS) for our photonic crystal. The 2 peaks are attributed to the 2 band edges of the photonic crystal. This ODOS is calculated by integrating the simulated band structure over the range of k-vectors that are expected to be collected through our measurement system (NA = 0.34). The observed features are in good agreement with the peak position of the long wavelength peak showing excellent agreement with the experimental spectra shown in Fig 5 a), which are the non-leaky modes shown in Fig 3.

V. SUMMARY

We have reported the realisation of an epitaxially regrown PCSEL at 1.5 μm, operating quasi-CW at room temperature opening a new route to surface emitting lasers in InP materials.

ACKNOWLEDGMENT

This work has been supported by the EPSRC Center for Doctoral Training in Photonic Integrations and Advanced Data Storage.

A.F.M gratefully acknowledges funding from the Royal Commission for the Exhibition of 1851 and CST Global Ltd.

R.J.E.T is funded by the RAEng Enterprise Fellowship.

REFERENCES

[1] K. Hirose et al., "Watt-class high-power, high-beam-quality photonic-crystal lasers" *Nature Photonics* **8**, 406-411, 2014.

[2] K. J.Reillya et al., "Epitaxial Regrowth and Hole Shape Engineering for Photonic Crystal Surface Emitting Lasers (PCSELs)"*Journal of Crystal Growth*, 2020.

[3] Y. Kurosaka et al., "Controlling vertical optical confinement in two-dimensional surface-emitting photonic-crystal lasers by shape of air holes", *Optics Express* **16** (22), 1848-18494, 2008.

[4] M. Imada et al., "Coherent two-dimensional lasing action in surface-emitting laser with triangular-lattice photonic crystal structure" *Applied Physics Letters* **75** (3), 316, 1999.

[5] K. Sakai et al., "Lasing band-edge identification for a surface-emitting photonic crystal laser", *IEEE Journal on Selected Areas of Communication* **23** (7), 1335, 2005.

[6] D. Ohnishi et al., "Room temperature continuous wave operation of a surface emitting two dimensional photonic crystal laser" *Optics Express* **12** (8), 1562, 2004.

[7] Y. Kurosaka et al., "Controlling vertical optical confinement in two-dimensional surface-emitting photonic-crystal lasers by shape of air holes*", Optics Express* **16** (22), 1848-18494, 2008.

[8] S. Noda et al., "Polarization Mode Control of Two-Dimensional Photonic Crystal Laser by Unit Cell Structure Design" *Science* **293**, 1123, 2001.

[9] R.J.E. Taylor et al., "Mode Control in Photonic Crystal Surface Emitting Lasers Through external reflection",
IEEE Journal of Selected Topics in Quantum Electronics 23, 6, 2017.

[10] Y. Kurosaka et al., "On-chip beam-steering photonic-crystal lasers", Nature Photonics, 4, 447 – 450, 2010.

[11] R. J. E. Taylor et al., "Coherently Coupled Photonic Crystal Surface Emitting Laser Array", *IEEE Journal of Selected Topics in Quantum Electronics* **21**, 2015.

[12] R. J. E. Taylor et al., "Electronic control of coherence in a two-dimensional array of photonic crystal surface emitting lasers", *Scientific Reports* **5**, 13203, 2015.

[13] R.J.E Taylor et al., "Band structure and waveguide modelling of epitaxially regrown photonic crystal surface emitting lasers" *Journal of Physics D* **46** (26) 264005(8pp), 2013.

[14] R.J.E. Taylor et al., "Optimisation of Photonic Crystal Coupling Through Waveguide Design", *Optical and Quantum Electronics* **49** (2), 47, 2017.

[15] G. Li et al., "Modelling and Device Simulation of Photonic Crystal Surface Emitting Lasers Based on Modal Index Analysis", *IEEE Journal of Selected Topics in Quantum Electronics* **99**, 1-1, 2019.

[16] S. G. Johnson and J. D. Joannopoulos, "Block-iterative frequency-domain methods for Maxwell's equations in a planewave basis" *Optics Express* **8** (3), 173-190, 2001.

[17] Y. Kurosaka et al., "Controlling vertical optical confinement in two-dimensional surface-emitting photonic-crystal lasers by shape of air holes", *Optics express* **16** (22), 1848-18494, 2008.

[18] Y. Kurosaka et al., "Band structure observation of 2D photonic crystal with various v shaped air hole arrangements" *IEICE Electronics express* **6** (13), 966, 2009.

[19] K. Sakai et al., "Coupled-wave model for square-lattice two-dimensional photonic crystal with transverse-electric-like mode", *Appl. Phys. Lett.* **89**, 021101, 2006.

[20] T. Inoue et al., "Design of photonic-crystal surface-emitting lasers with enhanced in-plane optical feedback for high-speed operation" *Optics Express* **28** (4), pp. 5050-5057, 2020.

[21] M. Nishimoto et al., "Air-Hole Retained Growth by Molecular Beam Epitaxy for Fabricating
GaAs-Based Photonic-Crystal Lasers" *Applied Physics Express* **6**, 042002, 2013.

[22] M. Nishimoto et al., "Fabrication of photonic crystal lasers by MBE airhole retained growth" *Applied Physics Express* **7**, 092703, 2014.

[23] D. M. Williams et al., "Epitaxially regrown GaAs-based photonic crystal surface emitting laser" *IEEE Photonics Technology* Letters **24** (11) 966-968, 2012.

[24] R. J. E. Taylor et al., "All-Semiconductor Photonic crystal surface emitting lasers based on epitaxial regrowth", *IEEE Journal of Selected Topics in Quantum Electronics* **19** (4), 4900407, 2013.

[25] M. Yoshida et al., "Fabrication of photonic crystal structures by tertiary-butly arsine-based metal-organic vapor-phase epitaxy for photonic crystal lasers", *Appl. Phys. Express* **9**, 2016.

[26] M. De Zoysa et al., "Photonic crystal lasers fabricated by MOVPE based on organic arsenic source", *IEEE Photon. Technol. Lett.*, **29**(20), 1739-1742, 2017.

[27] K. Ishizaki, M. De Zoysa, and S. Noda. "Progress in Photonic-Crystal Surface-Emitting Lasers." *Photonics*. Vol. 6. No. 3. Multidisciplinary Digital Publishing Institute, 2019.

[28] D. M. Williams et al., "Epitaxially regrown GaAs-based photonic crystal surface-emitting laser." *IEEE photonics technology letters* 24.11,966-968, 2012.

[29] M. Imada et al., "Multidirectionally distributed feedback photonic crystal lasers.'' *Physical Review B* 65.19, 195306, 2002.

[30] J.W. Demmel, "Applied Numerical Linear Algebra" 1st Edition, page 362-366, 1997.

[31] https://www.synopsys.com/photonic-solutions/rsoft-photonic-device-tools/passive-device-bandsolve.html.

[32] https://www.synopsys.com/photonic-solutions/rsoft-photonic-device-tools/active-device-lasermod.html.

[33] Y. Kurosaka et al., **"**Controlling vertical optical confinement in two-dimensional surface-emitting photonic crystal lasers by shape of air holes", *Optics Express* **16**(22),18458-18494, 2008.

[34] Y. Kurosaka et al., "Band structure observation of 2D photonic crystal with various Vshaped air hole arrangements", *IEICE Electronics Express* **6** (13), page 966-971, 2009.